\documentclass[aps,pra,twocolumn,superscriptaddress,letter,notitlepage]{revtex4-1}
\usepackage[english]{babel}
\usepackage{subeqnarray}
\usepackage{hyperref}
\usepackage{graphicx}
\usepackage{mathrsfs}
\usepackage{bm}
\usepackage{amssymb}
\usepackage{amsmath}
\usepackage{bbold}
\usepackage{xcolor}

\newcommand{\bra}[1]{\left<#1\right|}
\newcommand{\ket}[1]{\left|#1\right>}

\newcommand{\fig}[1]{Fig.~\ref{fig:#1}}
\newcommand{\abs}[1]{\left|#1\right|}
\begin{document}
\title{Atom-based coherent quantum-noise cancellation in optomechanics}
\date{\today}
\author{F. Bariani}
\affiliation{Department of Physics, College of Optical Sciences and B2 Institute, University of Arizona, Tucson, Arizona 85721, USA}
\author{H. Seok}
\affiliation{Department of Physics, College of Optical Sciences and B2 Institute, University of Arizona, Tucson, Arizona 85721, USA}
\affiliation{Department of Physics Education, Kongju National University, Gongju 314-701, South Korea}
\author{S. Singh}
\affiliation{ITAMP, Harvard-Smithsonian Center for Astrophysics, Cambridge, Massachusetts 02138, USA}
\author{M. Vengalattore}
\affiliation{Laboratory of Atomic and Solid State Physics, Cornell University, Ithaca, New York 14853, USA}
\author{P. Meystre}
\affiliation{Department of Physics, College of Optical Sciences and B2 Institute, University of Arizona, Tucson, Arizona 85721, USA}
\begin{abstract}
We analyze a quantum force sensor that uses coherent quantum noise cancellation (CQNC) to beat the Standard Quantum Limit (SQL). This sensor, which allows for the continuous, broad-band detection of feeble forces, is a hybrid dual-cavity system comprised of a mesoscopic mechanical resonator optically coupled to an ensemble of ultracold atoms. In contrast to the stringent constraints on dissipation typically associated with purely optical schemes of CQNC, the dissipation rate of the mechanical resonator only needs to be matched to the decoherence rate of the atomic ensemble -- a condition that is experimentally achievable even for the technologically relevant regime of low frequency mechanical resonators with large quality factors.  The modular nature of the system further allows the atomic ensemble to aid in the cooling of the mechanical resonator, thereby combining atom-mediated state preparation with sensing deep in the quantum regime. 
\end{abstract}
\maketitle

\section{Introduction}

The quest to measure forces at quantum limits of sensitivity \cite{braginsky1980, quant_meas} has been a primary thrust for the development of optomechanics \cite{OM_review} and the search for quantum effects in macroscopic mechanical resonators. For measurement schemes based on the optical detection of resonator motion, the trade-off between measurement imprecision due to shot noise and the measurement backaction on the resonator leads to a standard quantum limit (SQL). These competing effects have opposite scalings with the optical field intensity. Increasing the intensity to enhance the measurement strength and decrease shot noise results normally in increased measurement backaction, so that the route to improved force sensitivity requires therefore the mitigation or elimination of measurement backaction. 

Early proposals to achieve that goal relied on the insight that force sensitivity could be improved by measuring only a single quadrature of motion so as to isolate the effect of measurement backaction on the unseen orthogonal quadrature~\cite{thorne1978}. Such schemes are most conveniently realized by a temporal modulation of the electromagnetic field, as demonstrated in Ref.~\cite{hertzberg2010}. Generalizations of backaction-evading measurements to multimode systems have also been proposed~\cite{clerk2013}.

More recently a different approach to beyond-SQL measurements has been introduced. 
The idea is to introduce an `anti-noise' path in the dynamics of the optomechanical system via the addition of another oscillator that exhibits an equal and opposite response to the electromagnetic field, that is, an oscillator with a negative effective mass~\cite{HammererEPR}. Such an interaction realizes a coherent cancellation of measurement backaction via quantum interference~\cite{tsang2010,tsang2012}. 
An early proposal focused on the use of an ancilla cavity that is red-detuned from the optomechanical cavity, a quantum non-demolition coupling of the electromagnetic fields within the two cavities yielding the necessary anti-noise effect~\cite{tsang2010}. However, as shown in Ref.~\cite{CQNCPRA}, the technical requirements of this optical approach to CQNC pose prohibitive challenges, especially for low frequency mechanical resonators with large quality factors, the regime most relevant to force sensing applications. 

The coupling of optomechanical systems to the internal or external degrees of freedom of atomic ensembles has previously been shown to improve optomechanical cooling~\cite{treutlein_review,Genes2009,hammerer2010,Genes2011,treutlein2011,treutlein2013,bowen,geraci}, opening the possibility of ground state cooling outside the resolved sideband regime \cite{bariani2014}, and preliminary experimental results have confirmed the validity of that hybrid approach \cite{treutlein2014}. In addition, the interaction between mechanical resonators and atomic ensembles may be used to produce squeezed states, EPR-correlated states and entanglement \cite{HammererEPR}. This motivates considering hybrid systems that combine optomechanical resonators with the best features of atomic (or atom-like) systems to develop experimentally feasible  approaches to CQNC~\cite{HammererAnnalen}. A related approach has been experimentally shown to yield high sensitivity for coupled atomic oscillators \cite{Polzik2010, Polzik2011}. 

With these considerations in mind, this paper studies theoretically a CQNC scheme based on a dual cavity hybrid atom-optomechanical system where a mechanical oscillator used for force sensing is coupled to an ultracold atomic ensemble trapped in a separate cavity and serving as a negative mass oscillator.  This arrangement is a modification of the setup proposed for hybrid cooling~\cite{bariani2014}, the modular approach allowing to independently optimize the cavities for the mechanics and for the atoms. Importantly, we find that the interaction between the optomechanical cavity and the atomic ensemble can significantly alleviate the requirements for CQNC, and yields dramatic improvements in force sensitivity for parameters that have already been experimentally demonstrated. In addition, by rapidly modifying the external fields that control the atomic system it is possible to combine the spin-mediated optomechanical cooling of the mechanics with CQNC, thereby using the atomic ensemble both as a resource for state preparation and for force sensing deep into the quantum regime, 

The paper is organized as follows. Section II introduces the model of the system and derives the equation of motions for the relevant dynamical variables involved in the sensing process, which is discussed in detail in Section III. Section IV turns to force sensing, and computes the sensitivity limit of the atom-based CQNC scheme. Finally Section V is a summary and outlook. In Appendix, we show briefly how to combine the sensing scheme with hybrid cooling of the mechanics for realistic experimental parameters.

\section{Dual-cavity hybrid setup}
\label{model}
\begin{figure}
\includegraphics[width = \columnwidth]{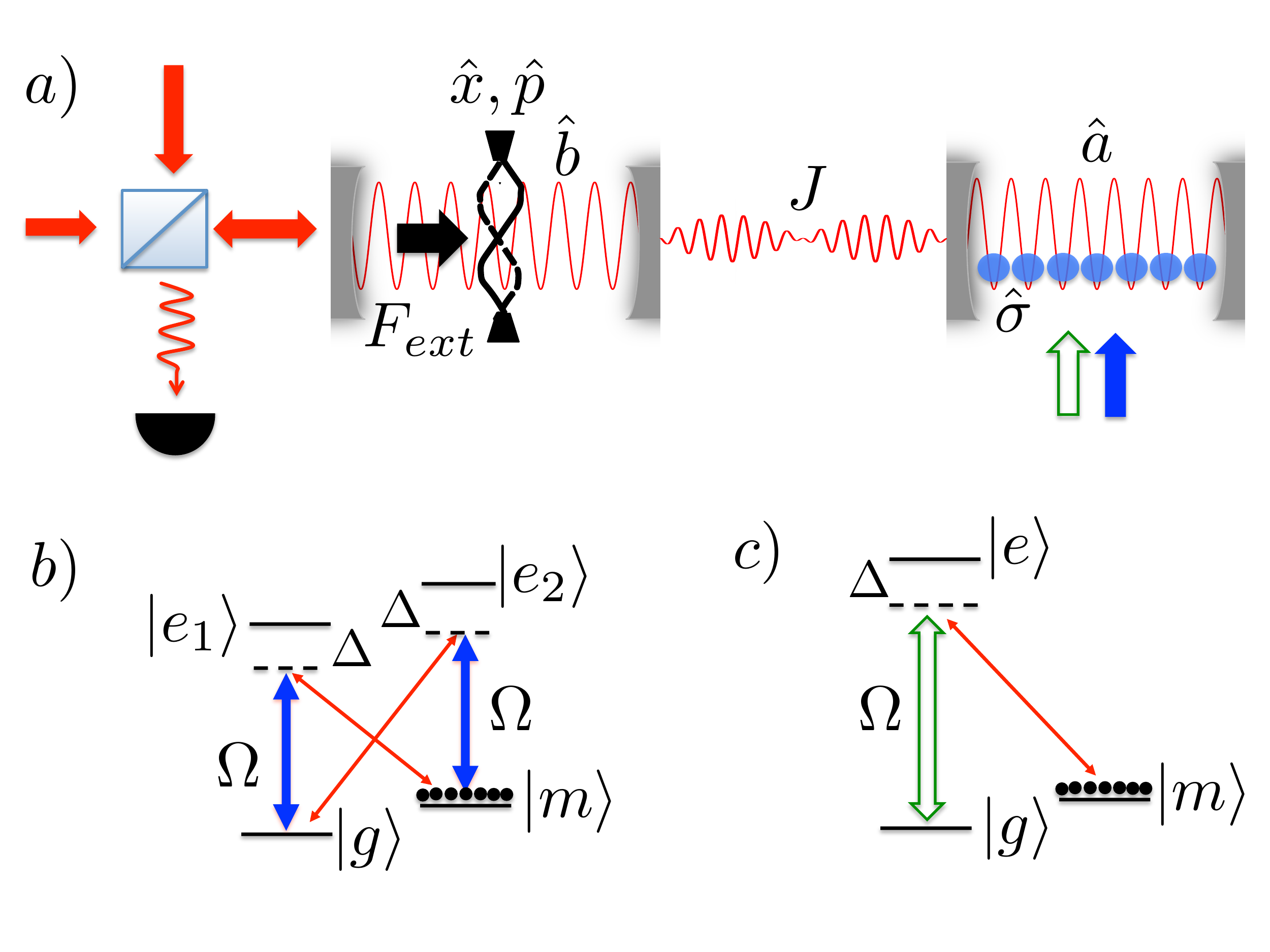}
\caption{(Colors online) (a) Sketch of the dual-cavity optomechanical-atomic hybrid sensing scheme, with two coherently coupled single-mode resonators, the left-hand side one containing the force sensing mechanics and the right-hand side one an ultracold atomic system that can be controlled by an external classical field. The blue (filled) arrow indicates the field used to generate an effective Faraday coupling of the atoms to the optical fields, see Section \ref{model}, and the green (empty) arrow is for the field used for Electromagnetically Induced Transparency (EIT) precooling of the mechanics, see Appendix \ref{precool}. (b) Atomic scheme leading to the effective Faraday interaction of Section \ref{model}, with a double $\Lambda$ atomic system driven by the intracavity field mode (thin red line) and a classical control field $\Omega$ (thick blue line) of frequency $\omega_{\Omega} = \omega_c$ resonant with the symmetric mode of the dual-cavity system. (c) Off-resonant Raman transition needed for EIT-based optomechanical precooling, resulting from the choice of control field frequency $\omega_{\Omega} = \omega_d + \omega_m$, see Appendix \ref{precool}.}
\label{fig:scheme}
\end{figure}

We consider a dual-cavity hybrid system, with one cavity -- the optomechanical resonator -- containing the mechanical oscillator acting as a force sensor, and the other -- the `atomic cavity'  -- containing a trapped ultracold atomic ensemble driven by the intracavity field and a classical control field. The two cavities are coupled coherently by their outcoupled fields, for example via an optical fiber link, see~\fig{scheme}.  

The total Hamiltonian comprises contributions from the intracavity fields, $H_{\rm field}$, the optomechanical interaction, $H_{\rm om}$, the atom-field coupling, $H_{\rm atom}$, the external driving of the cavities and dissipation, $H_{\rm res}$, and the forces to be measured, $H_{\rm ext}$,
\begin{equation}
H = H_{\rm field} + H_{\rm om} + H_{\rm at} + H_{\rm res}+ H_{\rm ext}.
\end{equation}

We take the optical cavities as single mode and model their coupling via a coherent tunneling term~\cite{HammererEPR,bariani2014}, so that 
\begin{equation}
H_{\rm field} = \hbar \omega_{\rm cav} (\hat a^\dagger \hat a+ \hat b^\dagger\hat b) +  \hbar J (\hat{b}^{\dagger}\hat{a} + \hat a^\dagger\hat b),
\end{equation}
where $\hat  a$ and $\hat b$ are the annihilation operators for the atomic and optomechanical cavity field mode, respectively, and we take their frequencies to be equal, $\omega_a = \omega_b = \omega_{\rm cav}$. The tunneling term $J$ is a function of the dissipation rates $\kappa_{a}$ and  $\kappa_b$ of these cavities. Its explicit form depends on the specific geometry~\cite{bariani2014}.

As will become more apparent in the next section it is useful to diagonalize $H_{\rm field}$ in terms of the symmetric and antisymmetric modes of the coupled cavity system
\begin{align}
&\hat{c} = \frac{1}{\sqrt{2}}(\hat{a} + \hat{b}), \\
&\hat{d} = \frac{1}{\sqrt{2}}(\hat{a} - \hat{b}). 
\end{align}
as
\begin{equation}
H_{\rm field} = (\omega_{\rm cav} + J) \hat{c}^{\dagger}\hat{c} + (\omega_{\rm cav} - J)  \hat{d}^{\dagger}\hat{d}.
\end{equation} 

The optomechanical Hamiltonian $H_{\rm om}$ describes the dynamics of a mode of vibration of the mechanics with effective mass $m$ and resonant frequency $\omega_m$. In the absence of external force acting on it, $F_{\rm ext}=0$, it is driven by the radiation pressure of the intracavity field $\hat{b}$,
\begin{eqnarray}
H_{\rm om} &=& \frac{\hbar \omega_{m}}{2} (\hat{x}^2 + \hat{p}^2) + g_0\hat{b}^{\dagger}\hat{b} \hat{x} \\
&=& \frac{\hbar \omega_{m}}{2} (\hat{x}^2 + \hat{p}^2) + \frac{g_0}{2} (\hat{c}^{\dagger}\hat{c} - \hat{c}^{\dagger}\hat{d} - \hat{d}^{\dagger}\hat{c} +\hat{d}^{\dagger}\hat{d})\hat{x}.\nonumber
\end{eqnarray}
Here $\hat{x}$ and $\hat{p}$ are the position and momentum of the mechanical mode, normalized to its zero point motion $x_{\rm zp} = \sqrt{\hbar/(m\omega_m)}$ and momentum $p_{\rm zp} = \hbar/x_{\rm zp}$, and $g_0$ is the single photon optomechanical coupling.

Turning now to the Hamiltonian $H_{\rm at}$, we consider an ensemble of $N$ ultracold atoms interacting non-resonantly with the intracavity field $\hat a$ and a classical control field of Rabi frequency $\Omega$ and frequency $\omega_\Omega$, see \fig{scheme}(b). We assume that the excited states $|e_1\rangle$ and $|e_2\rangle$ are sufficiently off-resonant that they can be adiabatically eliminated and the dynamics is restricted to two Raman-coupled levels $\ket{g}$ and $\ket{m}$ in the hyperfine ground state manifold. The light-atom coupling reduces then the coupled double-$\Lambda$ system to an effective two-state system driven by a Faraday interaction~\cite{HammererEPR, HammererRMP}. We further assume that a static external magnetic field tunes the Zeeman splitting between these states into resonance with the frequency $\omega_m$ of the mechanical resonator and that the atoms are initially pumped in the hyperfine level of higher energy, $\ket{m}$, resulting in an inverted ensemble that is well approximated for large $N$ by a harmonic oscillator of negative effective mass. 

Introducing the collective operator
\begin{equation}
\hat{\sigma} \equiv N^{-1/2}\sum_{j=1}^N \ket{m_j}\bra{g_j},
\end{equation}
where $j$ labels the different atoms, these approximations result in the effective Hamiltonian
\begin{equation}
H_{\rm at} =  -\hbar \omega_m \hat \sigma^\dagger\hat \sigma + \hbar G_0 \cos(\omega_\Omega t) (\hat a + \hat a^{\dagger}) (\hat \sigma + \hat \sigma^{\dagger}),
\end{equation}
or, in terms of the symmetric and antisymmetric modes, 
\begin{eqnarray}
H_{\rm at} &=& -\hbar \omega_m \hat \sigma^{\dagger}\hat{\sigma} + \frac{\hbar G_0}{\sqrt{2}} \cos(\omega_\Omega t) \nonumber \\
&\times& \left [  (\hat{c} + \hat{c}^{\dagger}) (\hat{\sigma} + \hat{\sigma}^{\dagger}) + (\hat{d} + \hat{d}^{\dagger}) (\hat{\sigma} + \hat{\sigma}^{\dagger}) \right ].
\label{eq:Hat}
\end{eqnarray}
Here 
\begin{equation}
G_0 = \sqrt{N} \mathcal{E} (\Omega/\Delta)
\end{equation}
is the collective Raman coupling between the atomic levels, with $\mathcal{E}$ being the cavity mode Rabi frequency, $\Omega$ the Rabi frequency of the control field, and $\Delta$ the detuning of the control beam from the excited atomic states. 
 
Finally, $H_{\rm res}$ accounts for the system interaction with its environment, including the dissipation of mechanics, cavities, and atoms as well as external pumping of the cavity modes with a laser of frequency $\omega_L$, and $H_{\rm ext}$ accounts for the coupling of the mechanics to the external forces to be measured.

\section{Heterodyne pumping and sensing}

The usual way to enhance the optomechanical interaction $g_0$ is to drive the optomechanical resonator with a classical light field. However, because the two resonators are coupled it is important to do so while preserving the polarization of the atomic ensemble, so that it still behaves as an effective harmonic oscillator of negative mass. One way to satisfy these apparently conflicting requirements is to exploit the  frequency splitting $2J$ between the symmetric and antisymmetric modes $\hat c$ and $\hat d$ of the coupled cavities system. Specifically, we pump the system at the frequency $\omega_d = \omega_{\rm cav} - J$  of the antisymmetric mode and measure the signal via a standard heterodyne detection scheme at the frequency $\omega_c = \omega_{\rm cav} + J$ of the symmetric mode~\cite{Keye2015}.  We further select which optical mode interacts with the atoms by adjusting the frequency $\omega_{\Omega}$ of the control field that sets the Raman two-photon resonance condition. It is easily verified that for $\omega_{\Omega} = \omega_c$ the second term in Eq.~(\ref{eq:Hat}) becomes resonant. Invoking in addition the rotating wave approximation we can ignore the small contribution due to the off-resonant third term in that equation, so that 
\begin{equation}
H_{\rm at} \approx -\hbar \omega_m \hat \sigma^\dagger \hat \sigma + \frac{G_0}{\sqrt{2}} (\hat c + \hat c^\dagger)(\hat \sigma + \hat \sigma^\dagger).
\end{equation}

The external pumping of the optomechanical resonator results in a small but finite displacement of the mechanics, which induces some degree of cross-talk between the modes $\hat{c}$ and $\hat{d}$. This leads in turn to a finite occupation of mode $\hat{c}$ which then drives some atomic population from the excited hyperfine level $|m\rangle$ to $|g\rangle$, resulting in a deterioration of the atomic polarization. This effect can be compensated  by applying  an external static force on the mechanical resonator to counterbalance the displacement due to radiation pressure. This force can be finely tuned by monitoring the atomic polarization via a fluorescence measurement of the population of state $\ket{g}$, which vanishes when the mechanical displacement is fully compensated.

While both the symmetric and antisymmetric mode  couple to the mechanics, the resulting sidebands at frequencies $\omega_{\rm cav} + J \pm \omega_m$ and $\omega_{\rm cav}  - J \pm \omega_m$ are well separated  and it is possible to filter out the small contribution of the off-resonant, anti-symmetric mode from the total Hamiltonian. Linearizing the resulting Hamiltonian around the steady state gives, in the frame rotating at the pumping frequency $\omega_d$,
\begin{eqnarray}
H_c &=& \frac{\hbar \omega_m}{2} (\hat{x}^2 + \hat{p}^2) - \frac{\hbar \omega_m}{2} (\hat{x}^2_{\sigma} + \hat{p}^2_{\sigma}) + \hbar J (\hat{x}^2_{c} + \hat{p}^2_{c}) \nonumber \\
&+&\hbar g \hat{x}_c \hat{x} + \hbar G \hat{x}_c \hat{x}_{\sigma}   + H'_{\rm ext},
\label{eq:Hc}
\end{eqnarray}
where
\begin{eqnarray}
\hat{x}_c &=& (\hat{c} + \hat{c}^{\dagger})/\sqrt{2}\,\,\,\,\,\,\,\,\,\,,\,\,\,\,\,\,\,\,\hat{p}_c = i(\hat{c}^{\dagger} - \hat{c})/\sqrt{2},\nonumber \\
\hat{x}_{\sigma}&=& (\hat{\sigma} + \hat{\sigma}^{\dagger})/\sqrt{2}\,\,\,\,\,\,\,\,,\,\,\,\,\,\,\,\,\hat{p}_{\sigma} = i(\hat{\sigma}^{\dagger} - \hat{\sigma})/\sqrt{2},\nonumber
\end{eqnarray}
$G = \sqrt{2} G_0$,  $g = - g_0 \langle\hat{d}\rangle/\sqrt{2}$, and $ H'_{\rm ext}$ is the same as $H_{\rm ext} $, but in a frame rotating at the frequency $\omega_d$.

The resulting Heisenberg-Langevin equations of motion are
\begin{align}
&\dot{\hat{x}} = \omega_m \hat{p},\\
&\dot{\hat{p}} = -\omega_m \hat{x} - g\hat{x}_c  - \gamma_m \hat{p} + \sqrt{\gamma_m} (\hat{f} + F_{\rm ext}),\\
&\dot{\hat{x}}_c = 2J \hat{p}_c - (\kappa/2) \hat{x}_c + \sqrt{\kappa} \hat{x}_c^{\rm in},\\
&\dot{\hat{p}}_c = - 2J \hat{x}_c - g \hat{x} - G \hat{x}_{\sigma} - (\kappa/2) \hat{p}_c +\sqrt{\kappa} \hat{p}_c^{\rm in},\\
&\dot{\hat{x}}_{\sigma} = -\omega_m \hat{p}_{\sigma} - (\Gamma/2) \hat{x}_{\sigma} + \sqrt{\Gamma} \hat{x}_{\sigma}^{\rm in},\\
&\dot{\hat{p}}_{\sigma} = \omega_m \hat{x}_{\sigma} - G \hat{x}_c -(\Gamma/2) \hat{p}_{\sigma} +\sqrt{\Gamma} \hat{p}_{\sigma}^{\rm in},
\label{HL}
\end{align}
where we have taken the two cavities to have the same decay rate, $\kappa_b = \kappa_a = \kappa$, with associated noise operators $\hat x_c^{\rm in}$ and $\hat p_c^{\rm in}$~\cite{walls}. The dissipation rate of the mechanics is $\gamma_m$, with noise operator $\hat f$, and the atomic coherence between the hyperfine states $|e\rangle$ and $|g\rangle$ decays at rate $\Gamma$, with noise operators $\hat x_\sigma^{\rm in}$ and $\hat p_\sigma^{\rm in}$. Finally $F_{\rm ext}$ is the (classical) external force to be measured. It  is normalized to $\sqrt{\hbar m \omega_m \gamma_m}$ and thus expressed in units of $\sqrt{\rm Hz}$. Since we are in the frame rotating at $\omega_d$, the sensing mode $\hat{c}$ appears to have an effective frequency $2J$.

These equations are similar to those describing the coherent quantum noise cancellation scheme proposed in Ref.~\cite{CQNCPRA}. As will be clear in the next section, in that latter case the role of the atoms is played instead by an ancilla cavity, and a crucial requirement is that this cavity be in the resolved sideband regime. The stringent requirement on the dissipation in the present scheme is passed onto the atoms: we have to match the mechanical dissipation with the rate of atomic decoherence, with $(\omega_m \gg \gamma_m = \Gamma)$, a condition that is relatively easy to satisfy \cite{quant_mem}. 

\section{Force Sensing}
\label{sensing}

Because of the dependence of the quadrature $\hat p_c$ on the position $\hat x$ of the mechanics, which is in turn coupled to its momentum $\hat p$ one can detect that external force by monitoring the symmetric mode $\hat c$, for instance via a hetrodyne measurement. We turn to the frequency domain in order to solve the dynamics of the relevant variables with respect to the input noise sources. We are interested in calculating $\hat{p}_c(\omega)$. Introducing the Fourier domain operators
\begin{equation}
\hat{O}(\omega) = \frac{1}{\sqrt{2 \pi}} \int dt \hat{O}(t) e^{- i \omega t},
\end{equation} 
Eqs.~(\ref{HL}) become
\begin{align}
&\hat{p} = (i\omega/\omega_m) \hat{x}, \\
&\hat{p}_{\sigma} = (\omega_m \hat{x}_{\sigma} + \sqrt{\Gamma} \hat{p}_{\sigma}^{\rm in} - G  \hat{x}_c )/(i\omega + \Gamma/2), \\
&{\hat{x}} = \chi_m [-g \hat{x}_c + \sqrt{\gamma_m} (\hat{f} + F_{\rm ext}) ], \label{eq:x}\\
&\hat{x}_{\sigma} = \chi_{\sigma}\{-G \hat{x}_c + \sqrt{\Gamma} [\hat{p}_{\sigma}^{in} - (i\omega + \Gamma/2) \hat{x}_{\sigma}^{\rm in}/\omega_m]\} \label{eq:xsigma}\\
&\hat{x}_c = \chi_c (2J \hat{p}_c + \sqrt{\kappa} \hat{x}_c^{\rm in}),\label{eq:xc}\\
&\hat{p}_c = \chi_c (- 2J \hat{x}_c - g \hat{x} - G \hat{x}_{\sigma} + \sqrt{\kappa} \hat{p}_c^{\rm in}),\label{eq:pc}
\end{align}
where we have introduced the effective susceptibilities
\begin{align}
&\chi_c  = \frac{1}{i\omega + \kappa/2}, \\
&\chi_m  = \frac{\omega_m}{\omega^2_m - \omega^2 + i \omega \gamma_m}, \\
&\chi_{\sigma} = \frac{-\omega_m}{\omega^2_m - \omega^2 + i \omega\Gamma + \Gamma^2/4} \label{eq:at_susc},
\end{align}
for the cavity field, the mechanics and the atomic ensemble, respectively.  Substituting Eqs. (\ref{eq:x})- (\ref{eq:xc}) into Eq.(\ref{eq:pc}), using the standard input-output relation~\cite{walls}
\begin{equation}
\hat{p}_c^{\rm out} = \sqrt{\kappa} \hat{p}_c - \hat{p}_c^{\rm in},
\end{equation}
and expressing the detected phase quadrature of the cavity field in terms of the input noise and signal we find
\begin{align}
p_c^{\rm out} =& \sqrt{\kappa} \chi'_c \{ - g \chi_m \sqrt{\gamma_m} (\hat{f} + F_{\rm ext})+ \nonumber \\
                  &+ \sqrt{\kappa} [ (1 - 1/(\chi'_c \kappa)) \hat{p}_c^{\rm in} - \chi_c 2J \hat{x}_c^{\rm in}] \nonumber\\
                  &- G \chi_{\sigma} \sqrt{\Gamma} [\hat{p}_{\sigma}^{\rm in} - (i\omega + \Gamma/2)\hat{x}_{\sigma}^{\rm in}/\omega_m] \nonumber \\
                  &+ \sqrt{\kappa} \chi_c [g^2 \chi_m + G^2 \chi_{\sigma}] \hat{x}_c^{\rm in} \},
\label{pout}
\end{align}
where we have introduced the modified quadrature susceptibility 
\begin{equation}
\frac{1}{\chi'_c} = \frac{1}{\chi_c} +2J\chi_c [2J - (g^2 \chi_m + G^2 \chi_{\sigma})]. 
\label{chi'}
\end{equation}
The first term in the detected quadrature (\ref{pout}) contains the signal and the thermal noise, the latter being independent of the detection frequency. The second line accounts for the shot noise of the field, modified by the term coupling the two field quadratures. The third line is the contribution of the atomic noise. Finally the last line describes the backaction on $\hat p_c$ of the conjugate variable $\hat x_c$.

The key point, and the essence of CQNC, is that for $g = G$ and $\chi_m = -\chi_{\sigma}$ the contributions to measurement backaction from the mechanics and from the atoms cancel each other for all frequencies. They can then be thought of as noise and antinoise contributions to the signal. This is the reason behind the need to create an oscillator with a negative effective mass in the atomic system. Note that these conditions also eliminate one of the terms from the susceptibility $\chi'_c$, see Eq.~(\ref{chi'}).  By matching the interaction strength of the field with both the mechanics and the atoms ($g = G$) and by arranging for the atomic coherence and the mechanical resonator to have the same dissipation rates ($\gamma_m = \Gamma$) one can therefore almost completely cancel the measurement backaction. In addition, if the mechanics has a quality factor high enough that  $\Gamma \ll \omega_m$  one can also eliminate the last term in the denominator of the atomic susceptibility (\ref{eq:at_susc}) to achieve perfect coherent backaction noise cancellation. 

To obtain the relationship between the force to be detected $F_{\rm ext}$ and the measured phase quadrature we rewrite Eq.~(\ref{pout}) as
\begin{equation}
F_{\rm ext} + \hat{F}_{\rm add} = \frac{-1}{2g\chi'_c \chi_m\sqrt{\gamma_m \kappa}} \hat{p}_{c}^{\rm out},
\end{equation}
this defines the added force noise $\hat{F}_{\rm add}$ as
\begin{align}
\hat{F}_{\rm add} = &\hat{f} - \sqrt{\frac{\kappa}{\gamma_m}} \frac{1}{g \chi_m} \left[ \left(1 - \frac{1}{\chi'_c \kappa}\right) \hat{p}_c^{in} - \chi_c 2J \hat{x}_c^{\rm in}\right] \nonumber \\
                      & + \frac{ G \chi_{\sigma} }{ g \chi_m }\sqrt{\frac{\Gamma}{\gamma_m}} \left[\hat{p}_{\sigma}^{\rm in} - \frac{i\omega + \Gamma/2}{\omega_m} \hat{x}_{\sigma}^{\rm in}\right] \nonumber \\
                      &  - \frac{g^2 \chi_m + G^2 \chi_{\sigma}}{g\chi_m}\sqrt{\frac{\Gamma}{\gamma_m}}\chi_c x_c^{\rm in}.
\end{align}
We assume that the quantum noise operators for both the symmetric mode for the optical field and the negative mass oscillator approximating the collective atomic coherence are well characterized by vacuum correlation functions, and that the mechanics is coupled to a thermal reservoir at temperature $T$ . From the general definition~\cite{CQNCPRA} of the noise spectrum of a random variable $F$
\begin{equation}
S_F(\omega)\delta(\omega-\omega') \equiv \frac{1}{2}  \left(\langle\hat{F}(\omega) \hat{F}(-\omega')\rangle + {\rm c.c.}\right),
\end{equation} 
we then find under the condition of exact backaction cancellation and for $\omega \ll \kappa$ that
\begin{widetext}
\begin{equation}
S_{F,\rm add}(\omega) = \frac{k_BT}{\hbar\omega_m}+\frac{1}{2} \left\{\frac{\kappa}{\gamma_m}\frac{1}{g^2\abs{\chi_m}^2} \left[\left(\frac{1}{2} - \frac{8J^2}{\kappa^2}\right)^2 + \frac{16J^2}{\kappa^2} \right] + \left(1 + \frac{\omega^2 + \Gamma^2/4}{\omega_m^2} \right)\right\}. 
\label{sfadd}
\end{equation}
\end{widetext}
This should be compared to the result of the standard optomechanical scenario~\cite{OM_review}
\begin{equation}
S_{F,\rm add}(\omega) =  \frac{k_BT}{\hbar\omega_m}  + \frac{1}{2} \left[\frac{\kappa}{\gamma_m}\frac{1}{g^2\abs{\chi_m}^2} \frac{1}{4} + 4g^2 \frac{1}{\kappa\gamma_m} \right], 
\end{equation}
with its familiar shot noise term scaling as $1/g^2$ and the backaction term scaling as $g^2$. In Eq.~(\ref{sfadd}), in contrast, the term proportional to $g^2$ is missing, indicating the successful elimination of measurement backaction. However, an immediate observation is that the cancellation of the back-action term occurs at the cost of an additional shot-noise-type term that is independent of the measurement strength $g^2$. 

The SQL, and its equivalent for the CQNC, are obtained by minimizing $S_{F, \rm add}$ with respect to $g$ at $T = 0$. This gives
\begin{equation}
S_{F, \rm SQL} = \frac{1}{\gamma_m \abs{\chi_m}},\\ 
\end{equation}
and
\begin{equation}
S_{F, \rm CQNC} = \frac{1}{2} \frac{\omega^2 + \omega_m^2 + \Gamma^2/4}{\omega_m^2}. 
\end{equation}
These noise spectral densities are dimensionless and must be multiplied by the normalization factor $\sqrt{\hbar m \omega_m \gamma_m}$ to recover values expressed in units of $\rm{N}^2/\rm{Hz}$. 

\begin{figure}[htbp]
\includegraphics[width = \columnwidth]{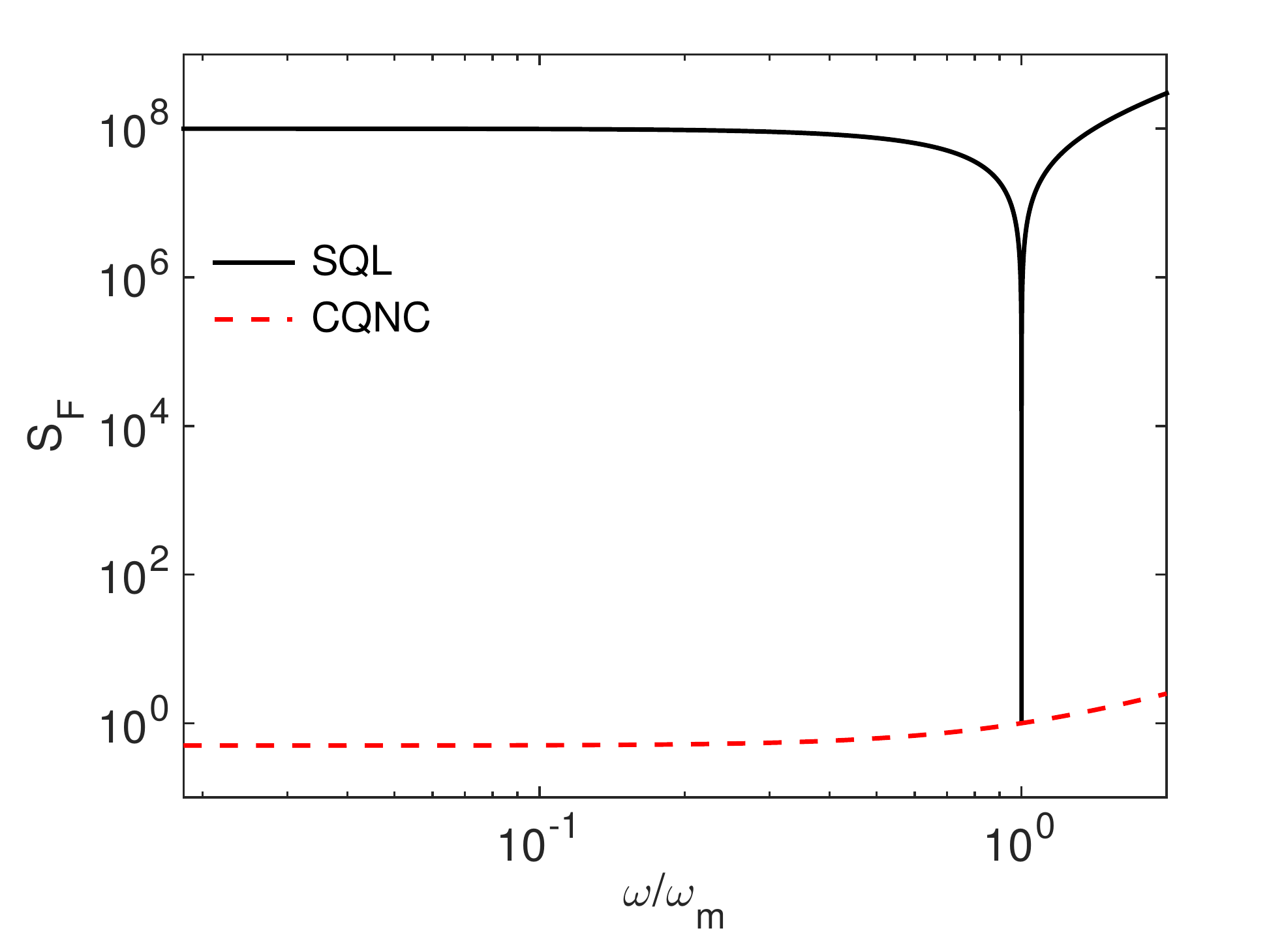}
\caption{(Colors online) Noise spectral density at optimal power for the standard optomechanical system (black solid line) and for the CQNC  scheme (red dashed line) as a function of the frequency $\omega$. Here $\omega_m = 2\pi \times 300$ kHz, $\kappa = 2\pi \times 1$ MHz and $Q = \omega_m/\gamma_m = 10^8$ \cite{chakram2014}. The spectral density is normalized to $\sqrt{\hbar m \omega_m \gamma_m}$. The sharp feature in the standard approach is the SQL achieved for a resonant driving force.}
\label{fig:SF}
\end{figure}

Figure~\ref{fig:SF} plots the behavior of the noise spectral density for both the normal and the CQNC situations. We see that the CQNC case offer shot-noise limited sensitivity over the whole detection bandwidth, and under optimum conditions provides a sensitivity equal to that of the conventional approach. From this point of view, then, the main advantage of CQNC is to enhance the bandwidth of detection.

\begin{figure}[htbp]
\includegraphics[width = \columnwidth]{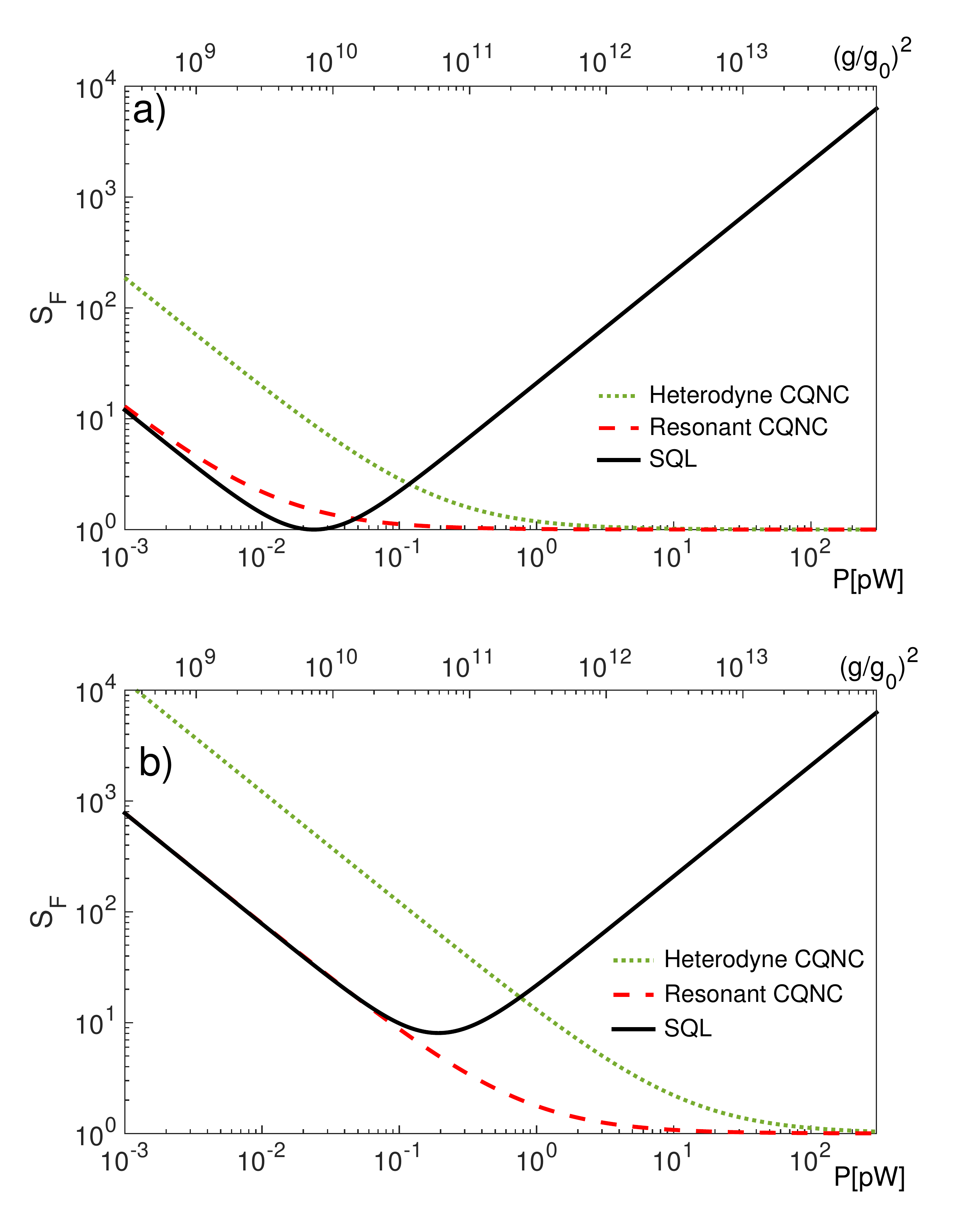}
\caption{(Colors online) Noise spectral density: (a) on mechanical resonance $\omega=\omega_m$ and (b) off-resonance (detuning of $4\gamma_m$), as a function of the  driving power $P = 2\hbar\omega_L\kappa(g/g_0)^2$ (bottom axis) for the standard optomechanical system (black solid), for the resonant CQNC (red dashed) and for the heterodyne CQNC (green dotted line) with $J = \kappa/\sqrt{2}$ for cavities arranged as in \cite{bariani2014}. We also display the dependence on the intracavity photon number $(g/g_0)^2$ on the top axis. Same parameters as in \fig{SF} and $\omega_L = 2\pi \times 384$ THz.}
\label{fig:SFcombined}
\end{figure}

Figure~\ref{fig:SFcombined} shows the noise spectral density as a function of the laser driving power, proportional to $g^2$, both on and off resonance. As expected, improved sensitivity over an increased bandwidth comes at the expense of requiring a higher driving power. Specifically, on resonance, the CQNC arrangement has the disadvantage of requiring a higher power to achieve the same sensitivity of the standard setup. However,  as we move away from the resonance $\omega=\omega_m$ the CQNC scheme rapidly becomes superior to the conventional approach, with no loss of sensitivity as the power is further increased. In this figure, we also compare the present case, indicated as \emph{heterodyne} CQNC, to the situation discussed in \cite{CQNCPRA}, called \emph{resonant} CQNC since pumping and sensing are done on the same mode.

In Appendix \ref{precool}, we show that it is possible for realistic parameters to use the atomic ensemble to {\it simultaneously} precool the vibrational mode to its ground state of motion and thus measure small forces on a mechanical resonator with minimal noise due to the suppression of the thermal component. 

\section{Conclusion}
In conclusion, we have presented the analysis of a coherent quantum noise cancellation (CQNC) scheme for beyond-SQL measurements in a hybrid quantum system that interfaces a cavity optomechanical system with an ensemble of ultracold atoms. We find that the condition for CQNC requires that the decoherence rate of the atomic ensemble is matched with the dissipation rate of the mechanical resonator -- a condition that is significantly more experimentally viable than the constraints on purely optical schemes for quantum noise cancellation. In addition, the modular nature of the hybrid quantum system considered here also allows for the atomic ensemble to be used as a quantum resource that enables enhanced optomechanical cooling, thereby combining atom-mediated state preparation with measurement of the mechanical resonator in the quantum regime. 

\acknowledgements
We thank E.S. Polzik, K. Hammerer, Y. S. Patil and S. Chakram for useful discussions. This work was supported by the DARPA QuASAR and ORCHID programs through grants from AFOSR and ARO, the NSF INSPIRE program, the U.S. Army Research Office, and the US NSF. M. V. acknowledges support from the Sloan Foundation. 

\appendix

\section{Precooling of the mechanics}
\label{precool}
In this Appendix, we adapt the formalism developed for hybrid optomechanical cooling of mechanics to the present system\cite{Genes2011,bariani2014}. 
We indicated in the main text that an advantage of the hybrid double-cavity system is the possibity to use the atomc sample both for precooling of the mechanics, thus eliminating thermal noise, and for CQNC sensing. Switching from one to the other configuration can be achieved simply by changing the classical control field $\Omega$. 

Precooling requires that the atoms be coupled to the optical fields in a single $\Lambda$ configuration, see~\fig{scheme}(c). In that case, the light tunneling between the cavities is filtered by the EIT susceptibility of the atoms maintained in the state $\ket{m}$ by cooling lasers. To achieve this goal in the present setup we pump the antisymmetric mode at frequency $\omega_d$ as before, but change the frequency of the control classical field $\omega_{\Omega} = (\omega_d + \omega_m)$ to achieve a Raman resonance with the antisymmetric mode. We neglect the effect of the symmetric mode since it is off-resonant and its effect on the cooling dynamics is strongly suppressed. 

Under these assumptions, the steady state population of mode $\hat{d}$ is 
\begin{equation}
\langle\hat{d}\rangle = -i \eta_d/(i\Delta_d + \kappa/2 - i \chi_{EIT}),
\end{equation}
where $\Delta_d = \omega_d - \omega_L$ is the detuning of the pumping laser of frequency $\omega_L$, and $\eta_d = \sqrt{P\kappa/(2\hbar\omega_L)}$ with $P$ the input laser power. The atomic  susceptibility is then $\chi_{\rm EIT} = -\mathcal{E}^2N/[\Delta + i \gamma_e/2 - \Omega^2/(\delta + i \Gamma/2)]$, where we have defined the single photon detuning $\Delta = \omega_L - \omega_{em}$, and the Raman detuning $\delta = \omega_L - \omega_d$. 

We can then evaluate the optical damping rate as difference between the effective cooling and heating rates as \cite{bariani2014}
\begin{eqnarray}\nonumber
\Gamma_{\rm opt}& = &\left(\frac{4G_0d}{ \sqrt{2} }\right)^2 
\mathrm{Re}\left[ \frac{1}{i(\Delta_d- \omega_m - \chi_{EIT,+}) + \kappa/2}  \right.\\
&&\left. - \frac{1}{-i(\Delta_d + \omega_m - \chi_{EIT,-}^{*}) + \kappa/2}\right]. 
\end{eqnarray}
Here, $\chi_{EIT,\pm}$ refers to the susceptibility evaluated at $\omega_L \pm \omega_m$. The final occupation number is then given by
\begin{equation}
n_{\rm min} = \frac{\gamma_m n_{\rm th} + \Gamma_{h}}{\gamma_m + \Gamma_{\rm opt}},
\end{equation}
where $n_{\rm th}$ is the thermal occupation of the relevant mode of the mechanical resonator.

To estimate $n_{\rm min}$ for a realistic hybrid optomechanical system, we assume the following parameters: $g_0 = 2\pi \times 300$ Hz, $\kappa = 2\pi \times 1$ MHz, $\omega_m = 2\pi \times 300$ kHz, $Q = \omega_m/\gamma_m = 10^8$ \cite{chakram2014}. We consider an atomic ensemble with $10^8$ atoms coupled to the cavity mode with strength $\mathcal{E} = 2\pi \times 100$ kHz and pumped with an external field of strength $\Omega = 50\gamma_e$, detuned from the single-photon transition by $\Delta = 50\gamma_e$. 
If we pump the cavity with a red detuning $\Delta_d = \omega_m$ and a power $P = 24 \mu$W, we obtain $\Gamma_{\rm opt} \approx 0.3 \omega_m$ and $n_{\rm min} < 1$ starting from room temperature $T = 300$K, with a lower bound given by $n_{\rm min} \approx 5 \times 10^{-2}$.

\end{document}